\documentclass[11pt,twoside]{article}
\usepackage{asp2010}

\def\not{{\sc not}}

\resetcounters

\begin{document}

\title{Pulsating hot subdwarfs with MS companions\\
or: EO\,Ceti is an sdO pulsator!}
\author{Roy~H.~{\O}stensen}
\affil{Instituut voor Sterrenkunde, K.U.~Leuven,
B-3001 Leuven, Belgium}

\begin{abstract}
About half of the hot subdwarfs are found to have spectra of composite
types, indicating a main sequence companion of spectral type F--K,
and the pulsators are no exception to this rule. The spectroscopic
contamination from the main sequence stars makes it hard to reliably
establish physical parameters for the hot component, and also makes
pulsations harder to detect as the amplitudes are depressed.
The binary fraction of the observed sample of hot subdwarf pulsators is
discussed, as are the biases that are affecting it.
Spectroscopic evidence is presented that clearly demonstrates that
the well known sdB pulsator, EO\,Ceti,
is misclassified, and is actually an sdOV star.
\end{abstract}

\section{Introduction}

Already when the discovery of pulsations in the hot subdwarf B (sdB) stars were
announced in the first papers by \citet{kilkenny97}, \citet{koen97} and
\citet{stobie97}, a strong binary fraction was evident, since all these
first pulsators are in composite systems.
At the time it was even speculated that the binarity is 
somehow involved in inducing the pulsations in these stars.
Much has happened since then. It is now clear that about 50\%\ of
sdB stars are found to be in close binaries with periods on the order of
hours or days \citep{copperwheat11}. The composite spectrum binaries are not
included among those.
The sdB stars with main sequence F-, G- or K-type companions
are known to form through stable Roche-lobe overflow rather than common
envelope ejection, and should end up with rather long periods. As recently
demonstrated from {\sc hermes} observations
(\O stensen \& Van Winckel, these proceedings), the periods of such binaries
are indeed exceedingly long, between one and three years.
It has been demonstrated by \citet{reed04}, by computing the IR excess of
the known subdwarfs in the field, that about half of sdB stars are such
composite spectrum binaries. Taken together, these results would seem to account
for $\sim$100\%\ of the sdBs. 

When the South African group 
(whose most prominent member, Dave Kilkenny, we are celebratinig the
65th birthday of with this conference) initiated the rush of
discoveries of pulsating sdB stars, the sample they observed was fairly
unbiased with respect to binarity. The only stars excluded would have been
those with main sequence companions of type A, and possibly the occasional
(sub-)giant K or M stars, where the sdB star would be practically invisible in
the optical. But when the collaboration that I have participated in joined
the fray with observations from the Nordic Optical Telescope
({\sc not}), we cheated a bit, using Uli Heber's extensive library of spectra
and model grids to select targets most likely to be pulsators.
Thus, we could increase our detection efficiency from 1.7\%\ to
$\sim$10\% \citep{sdbnot}. However, in the
process we automatically selected against composite spectrum binaries, as the
contamination makes the temperature determination very uncertain, especially in
low-res spectra with modest S/N. This fact was somewhat alleviated by the fact
that we pretty soon ran out of targets with well-determined physical parameters,
and by the time we started observing the SDSS sample we were less concerned
about the contamination.
Still, the consequence of these selection effects is that any attempt to check
if the fraction of binaries in the observed sample of sdB pulsators is the
same as in the population as a whole will be biased towards
non-composite systems. 
Revisiting the 49 stars compiled in Table~9 of \citet{sdbnot} reveals 
18 stars that are composite, i.e.~a fraction of 37\%, which is close enough to
the binary fraction of the field population to be consistent within 2-$\sigma$
given the sample size.  We list the 18 composites in Table~1.

\begin{table}[t]
\caption{Stars classified as sdB with a main sequence companion.
The column labeled 'Disc.' identifies the target roughly in order of discovery
(line number in Table~9 of \citealt{sdbnot}), and the group responsible for the discovery.
}\smallskip
\begin{center}\small
\begin{tabular}{llllll}\tableline\noalign{\smallskip}
Star name & Survey name & Class & $m_V$ & Disc. & References \\
\noalign{\smallskip}\tableline\noalign{\smallskip}
V361\,Hya & EC\,14026--2647 & sdB+G2 & 15.3 & 1/SA    & \citet{kilkenny97} \\
EO\,Ceti  & PB\,8783        & sdB+F0 & 12.3 & 2/SA    & \citet{koen97}, \\
          &                 &        &      &         & \citet{odonoghue98b} \\
UX\,Sex   & EC\,10228--0905 & sdB+G2 & 15.9 & 3/SA    & \citet{stobie97} \\
LM\,Dra   & PG\,1618+563B   & sdB+F2$^1$ & 13.5 & 15/\not & \citet{silvotti00} \\
V387\,Peg & HS\,2151+0857   & sdOB+IR$^2$ & 16.5 & 20/\not & \citet{ostensen01b} \\
EP\,Psc   & PG\,2303+019    & sdB+IR$^2$ & 16.2 & 24/\not & \citet{silvotti02a} \\
          & PG\,0048+091    & sdB+F5 & 14.3 & 28/SA & \citet{koen04} \\
          & PG\,0154+192    & sdB+G/K & 15.3 & 29/SA & \citet{koen04} \\
          & J1717+5805      & sdOB+G9 & 17.4 & 30/\not & \citet{solheim04} \\
          & PG\,1419+081    & sdOB+F/G & 14.9 & 32/\not & \citet{solheim06} \\
          & J1445+0002      & sdOB+G0 & 14.9 & 33/\not & \citet{solheim06} \\
          & J1642+4252      & sdOB+G0 & 14.9 & 34/\not & \citet{solheim06} \\
          & EC\,11583--2708 & sdB+F/G & 14.9 & 36/SA & \citet{kilkenny06} \\
          & PG\,1657+416    & sdOB+G5 & 15.9 & 39/\not & \citet{oreiro07} \\
          & JL\,166         & sdOB+F/G & 15.0 & 43/BB & \citet{barlow09} \\
          & HE\,2151--1001  & sdB+F/G & 15.6 & 46/\not & \citet{sdbnot} \\
          & PG\,1033+201    & sdB+F9 & 15.4 & 48/\not & \citet{sdbnot} \\
          & HE\,1450-0957   & sdB+F9 & 15.3 & 49/\not & \citet{sdbnot} \\
\noalign{\smallskip}\tableline
\end{tabular} \end{center} \small
{$^1$ The companion is a visual binary, not an actual close companion.}\\
{$^2$ The companion is identified based on IR-excess, not on spectroscopic features.}
\end{table}

The sdB+MS binaries are in general much more poorly studied than the rest of the sdBV sample.
The complication of having a strong spectroscopic companion means several things.
First of all the spectroscopic parameter determination becomes much more complicated and
regardless of the care taken in correcting for the contamination, the resulting spectroscopic
parameters will always be more uncertain. For asteroseismology there can be a significant
suppression effect in the photometric amplitudes, an effect that is also colour dependent,
since the cool companion suppresses the amplitudes in the red more than in the blue.
Note also that the companion may itself be pulsating;
\cite{ostensen11b} found that several of the composite systems observed with the {\em Kepler}
spacecraft shows photometric variability at the 1\%\ level with
periods on order of days or longer, consistent with gamma Dor
pulsations or rotational effects associated with spots.

I have long wanted to rectify the bias towards non-composite spectroscopic binaries
among the sdBVs in order to explore the possible difference in internal structure
in sdBs that form through different evolutionary channels, i.e.~common envelope
ejection, stable mass transfer or mergers. This was also the overarching topic of the 
Ph.D.~thesis of Maja \citet{MajaPhD}. During the sample selection process for the
long period binary study with {\sc hermes}
(\O stensen \&\ Van Winckel, these proceedings), I was hoping to include some sdBVs in
the sample, but as can be seen from Table~1, only EO\,Ceti is bright enough to be
observed with {\sc hermes}, and even it at $V$\,=\,12.3 is on the limit of what
can be done.
However, as I explored low-resolution spectra of EO\,Ceti and other sdB+F/G binaries
kindly made available to be by Betsy Green and Uli Heber, I found an unexpected
surprise, and the rest of this talk will be dedicated to this discovery.

\begin{figure*}[!ht]
\centering
\includegraphics[height=11.4cm,angle=-90]{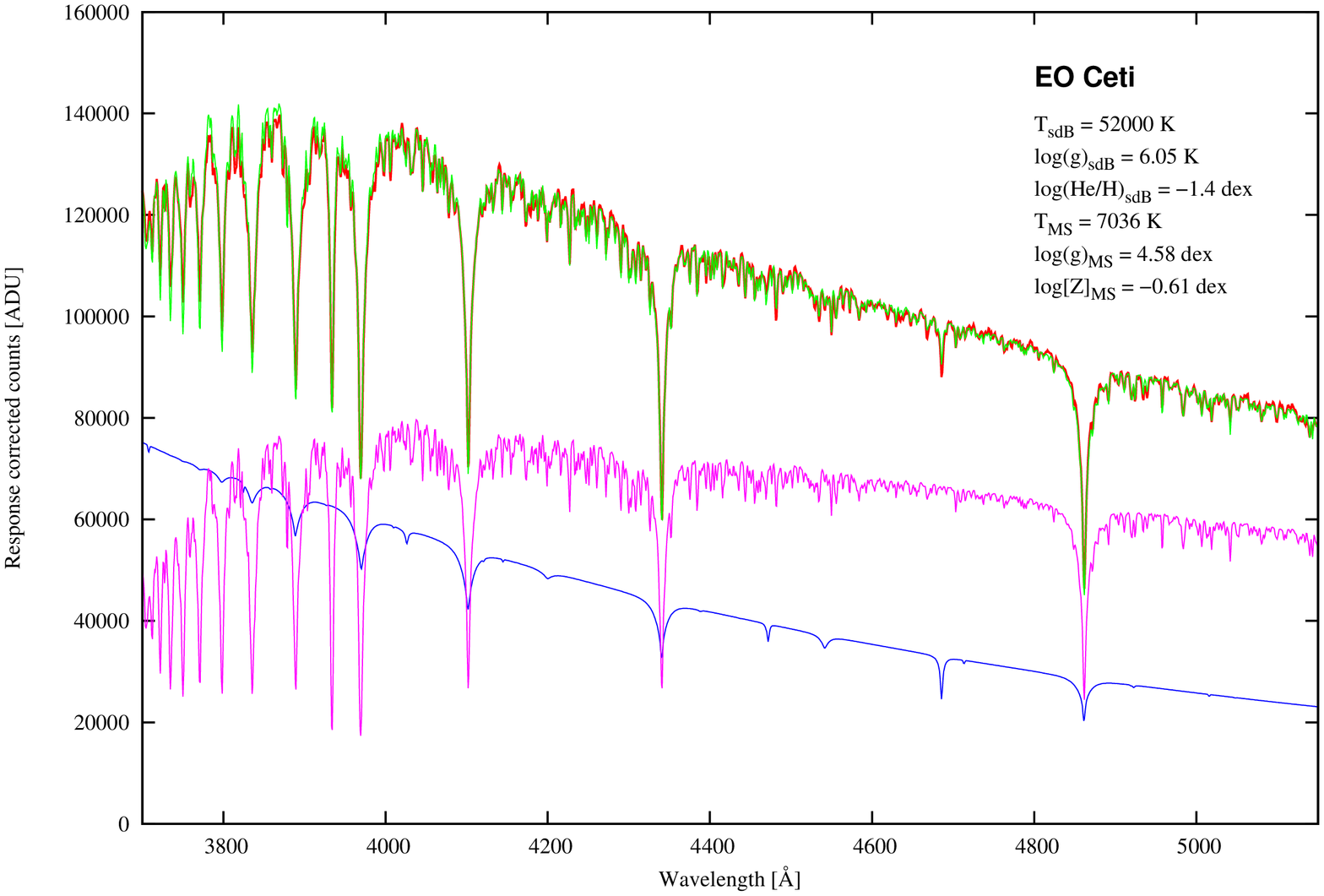}
\includegraphics[height=11.4cm,angle=-90]{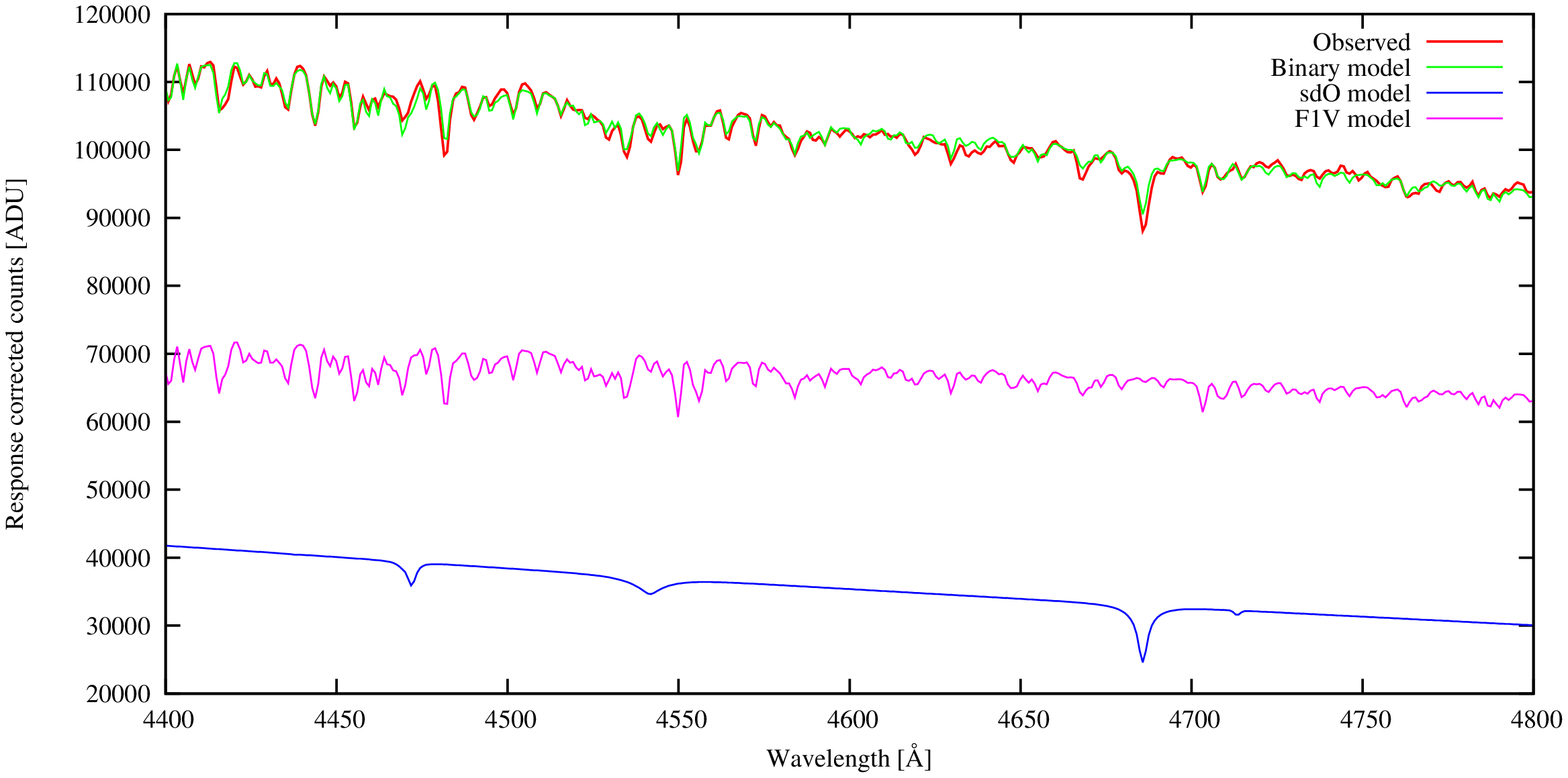}
\caption{
WHT/ISIS spectrum of EO\,Ceti.
Also shown are model spectra with parameters as stated on the plot.
The spectrum was response corrected by calibrating the instrumental response
function with a single sdB star.
The models have been scaled to fit the spectrum at 4050 and 5060\,\AA.
}
\end{figure*}

\section{Observations and analysis}

The spectrum of EO\,Ceti provided by Betsy Green showed no trace of any He\,{\sc i} lines
typical of sdB stars, but a line at 4686\,\AA\ looked as it might be He\,{\sc ii}.
If this was the case EO\,Ceti would not be the second sdBV ever found, but rather
the first sdOV. Pulsations in sdO stars were discovered much later than the sdBV
stars by \citet{woudt06}, and J16007+0748 remains to this date the only sdO pulsator
found. However, although this spectrum have exceptionally high S/N, the low resolution
(R\,$\approx$\,500) makes it difficult to identify individual lines among the strong
blends from the F-star companion. A high resolution {\sc feros} spectrum obtained
by Uli Heber confirmed the presence of a line at $\sim$4686\,\AA, but had insufficient
S/N to attempt a temperature determination. We therefore obtained 3 successive
deep integrations (900\,s) 
of EO\,Ceti with the {\sc isis} spectrograph on the William Herschel Telescope on
La Palma on August 27, 2010, using the R600B grating which provides a resolution
of R\,$\approx$\,2500 in the blue.  The spectra were extracted and summed, and
the response was corrected by calibrating with the spectrum of a hot DA white dwarf.
The resulting high S/N spectrum is shown in Figure~1. Also shown in the plot
is a simple binary model 
produced by scaling model spectra to fit at two well separated
points (here chosen to be 4050 and 5060\,\AA), after accounting for the
estimated reddening of the system, $E(B-V)$\,=\,0.04509 \citep{schlegel98}.
By subtracting one model from the spectrum, the parameters of the other
component could be fitted,
and this procedure was iterated until the procedure converged.
As can be seen from the figure,
this two-component solution makes an excellent fit to
practically every detail of the observed spectrum.
However, as can be seen in Figure~2, the solution can hardly be called unique.
For temperatures in the sdB--sdOB range (between 30\,000 and 45\,000\,K)
there is a strong
correlation between the temperature of the primary and that of the secondary,
whereas above $\sim$50\,000\,K changing the temperature of the primary does not
affect the $\chi^2$ significantly.
The procedure uses the whole available spectral range and is not sensitive
to the strength of the He\,{\sc i} and {\sc ii} lines.
In the bottom part of Figure~1 the zoom-in shows the region around the
He\,{\sc i} line 4472\,\AA\ and He\,{\sc ii} lines at 4512 and 4686\,\AA.
It appears that at the derived
primary temperature of 52\,000\,K, He\,{\sc i} at 4472 is still weaker in the observed
spectrum than in the model, while He\,{\sc ii} at 4686 appears to be stronger in the observed
spectrum. This indicates that the primary temperature could be significantly higher than
the derived solution, but not cooler. 
 
\begin{figure*}[t]
\centering
\includegraphics[height=\textwidth,angle=-90]{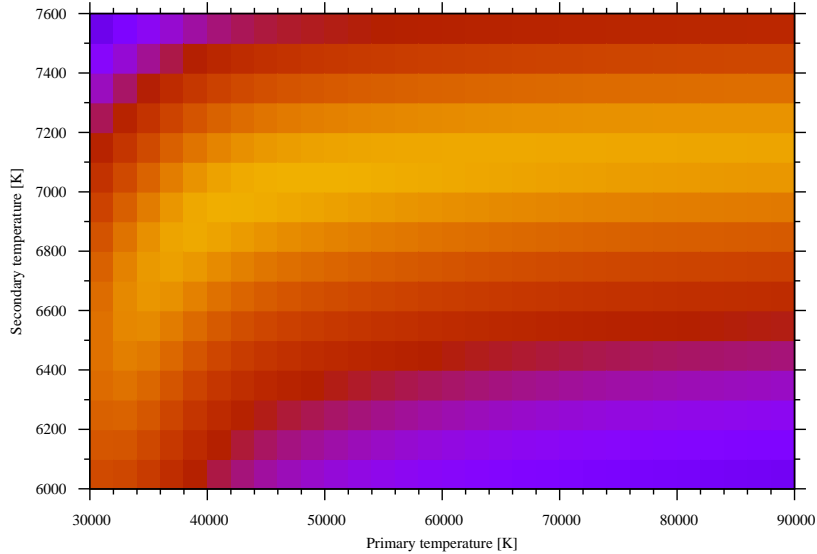}
\caption{
$\chi^2$ plane for model fits to the EO\,Ceti spectrum in Figure~1 as a function
of temperature of the two stars. Acceptable solutions to the overall shape of the
spectrum can be found for any temperature above $\sim$36\,000\,K.
Above $\sim$50\,000\,K the temperature of the secondary is no longer affected by
the choice of temperature for the primary.
}
\end{figure*}

We also obtained several spectra of EO\,Ceti with the {\sc hermes}
spectrograph \citep{raskin11} on the Mercator Telescope, also on La Palma.
These spectra have very high resolution (R\,$\approx$\,80\,000) but low S/N.
EO\,Ceti was observed on seven different occasions, using sequences of
three to four exposures between 1000 and 3600\,s each. The S/N in these
spectra are very low ($\sim$10) and plagued by cosmic rays, but a typical
S/N level of between 30 and 50 could be achieved by summing the individual
spectra and reducing the resolution by a factor of 4.
The final spectra were resampled to a uniform wavelength grid of 0.05\,\AA.
This is unproblematic as the rotational broadening of the MS component
is strong, and no sharp features (except a few interstellar lines) are present.
The full spectrum is to extensive to show here, but in Figure~3 we show
some of the most relevant regions, together with the same model spectra
as in Figure~2, convolved to the appropriate
resolution (0.05\,\AA\ and including a 72\,km/s rotational broadening
for the F-star).  This spectrum is a sum of three 2700\,s integrations
obtained on October 30, 2010.
As with the WHT spectrum, there is no trace of He\,{\sc i} at 4472\,\AA,
and He\,{\sc ii} at 4686\,\AA\ appears slightly stronger than in the model.
At this resolution it actually appears broader than the model rather than
deeper, which might well be the case as we have not made
any attempt at modelling the pulsational broadening of the lines.
The equivalent width of the line still points towards a primary
that is more likely to be hotter than the 52\,000\,K model than cooler.
The He\,{\sc i} line at 5876\,\AA\ might be present in the spectrum,
but is hardly significant.
In the lower panels of Figure~3 we show the model fit to some of the more
prominent lines from the F-star. The Ca\,{\sc ii} H and K lines are well
fitted, as are the Mg\,{\sc i}
triplet and the forest of Fe\,{\sc i} lines around 5400\,\AA.

\begin{figure*}[t]
\centering
\includegraphics[height=\textwidth,angle=-90]{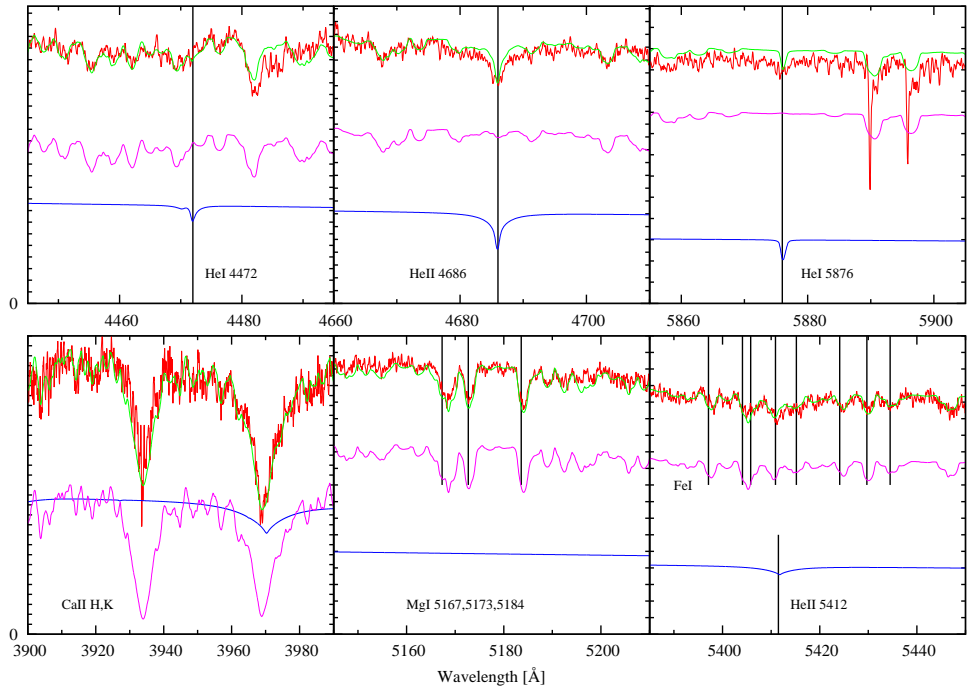}
\caption{
Selected lines from a
high-resolution spectrum of EO\,Ceti obtained with the {\sc hermes} spectrograph.
The model spectra shown are with the same parameters as in Figure~1.
}
\end{figure*}

\section{Conclusions}

After re-examination of new high quality spectroscopy it is
clear that EO Ceti is not an sdBV but an sdOV star.
Thus, as is the case with most other important discoveries in the
field of hot subdwarf stars, the first sdO pulsator was actually
discovered by Dave Kilkenny and his team, which I think is a very
appropriate discovery to announce at this conference held in his
honour.

The clear detection of the He\,{\sc ii}, together with weak or absent
He\,{\sc i} effectively forces EO\,Ceti to be classified as an sdO star,
regardless of its actual temperature. Using the slope of the spectrum
is no longer a viable option at temperatures in the sdO range, unless
one has access to UV spectroscopy. So far no attempt has been made to
reliably constrain the surface gravity of the component, and the
analysis does not force the two stars to be at the same distance.
However, since no orbit has been established, that may not actually
be the case. We will continue monitoring EO\,Ceti with high resolution
spectroscopy in the hope of detecting the orbital period of the system.
That may allow us to use spectral disentangling to decompose the
spectra into its two components, which will greatly simplify the
parameter estimation. However, the broadening from the pulsational
velocity at the surface of the sdO star and the broadening due to the
rotational velocity of the F star are both higher than the expected
orbital velocity, which makes this a very challenging project.

At B=12.3 EO\,Ceti is a much more encouraging target to study than
J16007+0748, since high-precision photometry can be obtained with
modest telescopes, and time-resolved spectroscopy is a lot more feasible.
In fact, thanks to its early discovery, EO\,Ceti is one of the
best studied hot subdwarf pulsators with extensive campaign
observations by \citet{odonoghue98b}, multicolor photometry by
\citet{vuckovic10} and an early attempt at time-resolved photometry
by \citet{jeffery00}. 

Can there be other sdO pulsators in the sample of known sdBVs
(Table 9 of \citealt{sdbnot})? Perhaps.
I have checked the ones that I have spectroscopy on, but no clear
candidates have emerged. But there are quite a few in that table
that have very poor spectroscopy, and many have no established helium
abundance. The spectroscopy also suffer from having been obtained with
many different instruments and various resolutions and signal levels.
In many cases only the original classification spectrum exists, so
clearly, much more work has yet to be done.

\acknowledgements
The research leading to these results has received funding from the European
Research Council under the European Community's Seventh Framework Programme
(FP7/2007--2013)/ERC grant agreement N$^{\underline{\mathrm o}}$\,227224
({\sc prosperity}), as well as from the Research Council of K.U.Leuven grant
agreement GOA/2008/04.

Based on observations obtained with the HERMES spectrograph, which is supported
by the Fund for Scientific Research of Flanders (FWO), Belgium,
the Research Council of K.U.Leuven, Belgium,
the Fonds National Recherches Scientific (FNRS), Belgium,
the Royal Observatory of Belgium, the Observatoire de Gen\'eve,
Switzerland and the Thüringer Landessternwarte Tautenburg, Germany.

\bibliographystyle{asp2010}
\bibliography{sdbrefs}

\end{document}